%% file: main.tex
\PassOptionsToPackage{table,xcdraw}{xcolor}
\documentclass[sigconf,authorversion]{acmart}
\usepackage[colorinlistoftodos]{todonotes}
\usepackage{multirow}
\usepackage{array}
\newcolumntype{M}[1]{>{\centering\arraybackslash}m{#1}}
\usepackage[utf8]{inputenc}
\usepackage[table]{xcolor}

\AtBeginDocument{%
  \providecommand\BibTeX{{%
    \normalfont B\kern-0.5em{\scshape i\kern-0.25em b}\kern-0.8em\TeX}}}

\copyrightyear{2021}
\acmYear{2021}
\setcopyright{acmcopyright}\acmConference[SIGCSE '21]{Proceedings of the 52nd ACM Technical Symposium on Computer Science Education}{March 13--20, 2021}{Virtual Event, USA}
\acmBooktitle{Proceedings of the 52nd ACM Technical Symposium on Computer Science Education (SIGCSE '21), March 13--20, 2021, Virtual Event, USA}
\acmPrice{15.00}
\acmDOI{10.1145/3408877.3432417}
\acmISBN{978-1-4503-8062-1/21/03}

\begin{document}

\title[Inquiry-Based Conceptual Feedback vs. Traditional Detailed Feedback in Software Testing Education: An Empirical Study]{A Comparison of Inquiry-Based Conceptual Feedback vs. Traditional Detailed Feedback Mechanisms in Software Testing Education: An Empirical Investigation }

\author{Lucas Cordova}
\affiliation{%
 \institution{Western Oregon University}
 \city{Monmouth}
 \state{OR}
 \country{USA}
}
\email{cordoval@wou.edu}

\author{Jeffrey Carver, Noah Gershmel}
\affiliation{%
 \institution{University of Alabama}
 \city{Tuscaloosa}
 \state{AL}
 \country{USA}
}
\email{carver@cs.ua.edu}
\email{nggershmel@crimson.ua.edu}

\author{Gursimran Walia}
\affiliation{%
 \institution{Georgia Southern University}
 \city{Statesboro}
 \state{GA}
 \country{USA}
}
\email{gwalia@georgiasouthern.edu}


\renewcommand{\shortauthors}{}

\begin{abstract}
The feedback provided by current testing education tools about the deficiencies in a student’s test suite either mimics industry code coverage tools or lists specific instructor test cases that are missing from the student’s test suite.  
While useful in some sense, these types of feedback are akin to revealing the solution to the problem, which can inadvertently encourage students to pursue a trial-and-error approach to testing, rather than using a more systematic approach that encourages learning. 
In addition to not teaching students why their test suite is inadequate, this type of feedback may motivate students to become dependent on the feedback rather than thinking for themselves. 
To address this deficiency, there is an opportunity to investigate alternative feedback mechanisms that include a positive reinforcement of testing concepts. 
We argue that using an inquiry-based learning approach is better than simply providing the answers. 
To facilitate this type of learning, we present Testing Tutor, a web-based assignment submission platform that supports different levels of testing pedagogy via a customizable feedback engine. 
We evaluated the impact of the different types of feedback through an empirical study in two sophomore-level courses. 
We use Testing Tutor to provide students with different types of feedback, either traditional detailed code coverage feedback or inquiry-based learning conceptual feedback, and compare the effects. 
The results show that students that receive conceptual feedback had higher code coverage (by different measures), fewer redundant test cases, and higher programming grades than the students who receive traditional code coverage feedback.
\end{abstract}



\begin{teaserfigure}
\end{teaserfigure}

\maketitle

\section{Introduction}
In programming courses, students often write and debug code by trial-and-error, running it on sample input (often provided by an instructor), or just using the compiler (i.e. believing that if it compiles, it must be correct)~\cite{edwards_2004}. 
Students’ lack of testing knowledge and ability mistakenly leads them to believe that they can determine the correctness of their code from a small number of test cases. 
They do not foresee their programs failing other test cases that may be used by their instructor and thus do not understand why they receive a low grade. 
Moreover, after their formal education, students are not properly equipped to enter the workplace because many graduate with a knowledge gap about software testing~\cite{begel_simon_2008,haddad_2002,kitchenham_budgen_brereton_woodall_2005,lethbridge_1998,lethbridge_2000,lethbridge_2000_1,wang_2008}

Researchers have developed pedagogical tools~\cite{collofello_vehathiri_2005,spacco_hovemeyer_pugh_emad_hollingsworth_padua-perez_2006, edwards_2003} to address the shortcomings in testing education. 
These tools (e.g., Marmoset, WebCAT) each provide students with different types of information about their code and tests including information like: whether the instructor’s tests have passed, code coverage, and even pointing out exactly which portion of code is problematic.
While these tools do provide useful information, one of the main drawbacks is that these tools tend to provide the ‘answers’ (usually after the students meet some criteria). 
For example, a tool might tell a student exactly which of the instructor’s test cases are missing from his or her test suite. 
In addition, code coverage tools indicate exactly which portions of code are not fully tested. 
While this information may be useful for improving the current test suite, beginning students lack the metacognitive ability to use that information to identify their knowledge gaps about fundamental testing concepts and determine why their test suite was inadequate. 
Tools that provide this type of automated feedback may actually discourage students from thinking on their own and instead encourage them to rely upon the automated feedback~\cite{buffardi_2015}.

We approach this limitation by focusing on an inquiry-based learning (IBL) approach~\cite{prince_felder_2006}. 
IBL is based on the Cognitive Constructivist learning theory~\cite{vygotskij_cole_1981} that students are more engaged when they actively construct and validate their own knowledge (through experience) rather than simply receiving the answers. 
IBL has improved student learning in many disciplines including: Social Science~\cite{justice_warry_cuneo_2002}, Geography~\cite{spronken-smith_angelo_matthews_o'steen_robertson_2007}, Psychology~\cite{muukkonen_lakkala_hakkarainen_2005}, Medicine~\cite{houlden_raja_collier_clark_waugh_2004}, Physics~\cite{abell_2005}, Meteorology~\cite{yarger_gallus_taber_boysen_castleberry_2000}, Chemistry~\cite{zoller_1999}, and Forestry~\cite{Yin2006PreparingRA}. 
Meta-analyses of IBL experiments showed that, compared with a traditional classroom approach, IBL results in an improved academic achievement, deeper understanding of content, critical thinking skills, motivation, engagement, and creativity~\cite{shymansky_hedges_woodworth_1990,smith_1996}. 
Previous work about IBL in CS (when used to teach the LOGO programming language) has shown that learning should emphasize discovery and that there needs to be some tutor facilitation to guide the learning process rather than leaving it open-ended~\cite{pettit_2015,hu_kussmaul_kneable_mayfield_yadav_2016}.

To implement an IBL-based approach for improving software testing education, we built \textbf{Testing Tutor}, a web-based assignment submission platform that supports testing pedagogy via a customizable feedback engine that can be integrated into any level of the CS curriculum.  
Testing Tutor’s innovative pedagogical contribution is in the type of feedback it provides. 
Rather than providing a student with the 'answer' (e.g. exactly which tests are missing) or the level of test coverage, Testing Tutor provides students with \textit{conceptual feedback}. 
\textit{Conceptual feedback} informs the student which underlying fundamental testing concepts their test suites does not adequately cover and provides suggestions for the student to initiate their own learning process about those concepts. 
This type of feedback will allow the student to determine on his or her own how to improve their test suite, rather than being told by the system exactly what is missing. 
Examples of \textit{conceptual} feedback for a CS2-level assignment are “the test suite has not fully tested all boundary conditions” and “the test suite misses part of a compound Boolean expression” along with resources provided (examples, videos). 

The primary goal of this paper is 
\textbf{to assess whether \textit{conceptual} feedback helps students produce better, more concise and comprehensive test suites.}

\section{Previous Approaches}
\label{sec:Previous}

Some educators have developed approaches to improve software testing pedagogy. 
Testing Tutor builds upon the ideas and shortcomings in these existing approaches. 

Collofello and Vehathiri~\cite{collofello_vehathiri_2005} developed a testing simulator that executes student test cases against built-in buggy programs and displays missing test cases. 
The tool reports the following metrics: 1) \textit{test completeness} – a measure of input coverage, 2) \textit{flow coverage} – a measure of statement/path coverage, 3) \textit{correctness} – a measure of student test outputs correspondence to expected test outputs, and 4) a \textit{fault detection metric} – a measure of fault detection effectiveness. 
At the conclusion of the testing exercise, the tool gives the students the ’answers’ (e.g. the correct set of tests). 
While this approach does encourage students to improve the test suite for their current project, it does not provide feedback about why the test suite is incomplete or help students carry the knowledge over to future assignments. 
Testing Tutor borrows the idea of learning how to test by testing someone else’s code. 
But different from this approach, which allows students to succeed through trial-and-error, Testing Tutor provides \textit{conceptual} feedback about why the test suite is incomplete and provides support for learning fundamental testing concepts that should carry over to future assignments.

Marmoset~\cite{spacco_hovemeyer_pugh_emad_hollingsworth_padua-perez_2006} and WebCAT~\cite{edwards_2004,edwards_2003,edwards_2003_1} use test coverage to provide students with automated feedback on their code. 
Marmoset provides public tests (i.e. visible to students) and private tests (i.e. not visible to students). 
When students’ code passes the public tests, Marmoset executes the private tests and informs the student of how many private tests failed along with the names of two of those failed tests. 
Similarly, WebCAT uses a concept of public and private tests to perform automated grading and provides students with feedback during the development process. 
WebCAT provides students with detailed feedback on failed tests and can provide code annotations with suggestions for improvement. 
A major drawback to these tools is the extra grading of tests cases required of the instructors. 
Testing Tutor borrows the idea of encouraging students to write better test cases but does not add additional work for instructors to grade test cases. 
In addition, rather than providing feedback about specific test cases, Testing Tutor provides guidance on how students can improve their own tests, which should encourage students to better learn the underlying testing concepts.

ProgTest~\cite{souza_2014} uses two sets of code and test cases, one provided by the student and one provided by the tool or the instructor. 
ProgTest runs the student's tests against both their own code and the code provided by the tool and runs the tests provided by the tool against the student's code. 
By comparing the coverage between the student's code and the instructor's code, the student can continuously improve their code and tests.
This approach produces higher quality outcomes in terms of semantic correctness compared with developing tests only after developing a complete code solution~\cite{kazerouni_2019}. 
Testing Tutor expands upon the type of feedback provided by ProgTest to include \textit{conceptual} information about the tests that are missing.

\section{Testing Tutor}
\label{sec:TestingTutor}
Testing Tutor is a web-based tool that supports testing pedagogy by helping students learn to develop higher-quality test suites and become more effective testers. 
Testing Tutor uses a reference implementation and its corresponding test suite to identify which tests are missing from a student’s test suite. 
Each missing test has one or more fundamental testing concept (e.g. testing boundary conditions or testing for data integrity) attributed to it. 
Based on the missing tests, Testing Tutor helps the students understand which fundamental testing concept knowledge they lack so they can improve their own test suite. 
Testing Tutor can be used in either \textit{Learning Mode} or \textit{Development Mode}. 
In Learning Mode, Testing Tutor teaches a student how to develop a complete test suite for a reference implementation of a program. 
In Development Mode, Testing Tutor helps a student completely test their own newly written code. 
Learning Mode can be used by itself as a testing exercise or to prepare students to write their own code for later use with  Development Mode. 
If instructors use Learning Mode in a stand-alone fashion (i.e. students will not later implement the assignment), then instructors can decide whether students see the reference code (white-box testing) or only see the specification (black-box testing). 
If instructors use Learning Mode to prepare students for the Development Mode, then the students will not see the reference code (black-box testing).

Testing Tutor is different from existing software testing education pedagogy because of its’ customizable feedback mechanism. 
Testing Tutor allows an instructor to tailor the level and type of feedback provided to the students. 
By annotating the test cases in the reference implementation and configuring the type of feedback Testing Tutor provides, an instructor can choose which learning concepts she or he wants to be the focus. 
Currently, Testing Tutor supports three types of feedback mechanisms, as discussed in this paper (\textit{detailed} feedback, \textit{conceptual} feedback, and no feedback). 
This design creates an opportunity for instructors and researchers to investigate which feedback mechanisms best promote learning and improvement while teaching software testing concepts. 

The features of Testing Tutor include:
\begin{itemize}
    \item \textit{Web-based interface} - access to the tool via any web browser;
    \item \textit{Authorization and authentication management} - supports institution hierarchies for courses, students, faculty, administrators, and assignments;
    \item \textit{Multi-institution support} - each institution can configure Testing Tutor so that users, assignments, and reports remain separate from other institutions;
    \item \textit{Assignment repository} - assignments can be private (to the instructor), shared only within an institution, or public;  
    \item \textit{Tailored feedback} - instructors select the type of feedback: no feedback, \textit{detailed} feedback, or \textit{conceptual} feedback;
    \item \textit{Course management} - generate reports for courses, instructors, and individual students, and analysis for assignments or groups of assignments; and
    \item \textit{Plug-in platform architecture} - facilitates the addition of more programming languages and is built to scale.
\end{itemize}

\section{Experiment}
\label{sec:Experiment}
To investigate the impact of \textit{conceptual} feedback compared with traditional \textit{detailed} code coverage feedback using Testing Tutor's \textit{Learning Mode}, we pose the following high-level research questions:

\vspace{8pt}
\noindent
\textit{\textbf{RQ1:} How do different types of feedback (\textit{conceptual}, \textit{detailed}, none) affect the quality of student test suites?}

\vspace{8pt}
Then, to specifically evaluate the usefulness of Testing Tutor, we pose a second research question:

\vspace{8pt}
\noindent
\textit{\textbf{RQ2:} What are the students' perception of the usefulness of Testing Tutor in terms of its usability and the feedback provided?}

To answer these research questions, we conducted a series of two quasi-experiments (an initial study followed by a replication).
The remainder of this section provides details on these studies.

\subsection{Participating Subjects and Artifacts}
We performed the studies in a sophomore-level software testing course at Oregon Institute of Technology in the Spring and Summer 2019 semesters.
We chose this course because it has the goal of helping students produce and improve the quality of code. 
Prior to this course, students completed CS1, CS2 and CS3 (Data Structures). 

For each study (the original and the replication), we split the students into two groups based on their course section. 
The students in Group A received the traditional \textit{detailed} coverage feedback.
The students in Group B received the \textit{conceptual} feedback.
Table~\ref{tab:compositions} illustrates the number of participants in each group. 

\begin{table}[!htb]
  \caption{Study group compositions}
  \label{tab:compositions}
  \begin{tabular}{ccc}
    \toprule
    Study & Group A Participants & Group B Participants\\
    \midrule
    Spring 2019&15&16\\
    Summer 2019&13&15\\
  \bottomrule
\end{tabular}
\end{table}

Over the course of the study, the students in each group received the same five assignments, all written in Java 1.11.  
We instructed the students to produce the most comprehensive, yet smallest, test suite possible for each assignment. 
The five assignments were:
\begin{itemize}
    \item \textit{Assignment 1} - An I/O program (a calendar program taking a date as input and returning the date of the day before, the day after, one week before, or one week ahead).
    \item \textit{Assignment 2} - A state-based data structure (a queue) supporting all queue operations and exception-handling.
    \item \textit{Assignment 3} - An object-oriented calculator containing multiple interfaces and inheritance.
    \item \textit{Assignment 4} - A comma-separated value (CSV) parser built using the Visitor design pattern.
    \item \textit{Assignment 5} - A banking application built using the Observer pattern for support of multiple clients.
\end{itemize}

\subsection{Independent Variable -- \textit{Type of Feedback}}
To measure the impact the \textit{type of feedback} had students’ testing performance, we defined the following levels for the independent variable, and configured Testing Tutor accordingly:
\begin{itemize}
    \item \textit{Treatment A} - Traditional \textit{detailed} feedback similar to code coverage output from tools like JaCoCo and CodeCover.
    \item \textit{Treatment B} - \textit{Conceptual} feedback which provides the student with the testing concepts that are not adequately tested from the perspective of the instructor and includes resources to review (textual and video).
\end{itemize}

\begin{figure}[!htb]
  \centering
  \includegraphics[width=\linewidth]{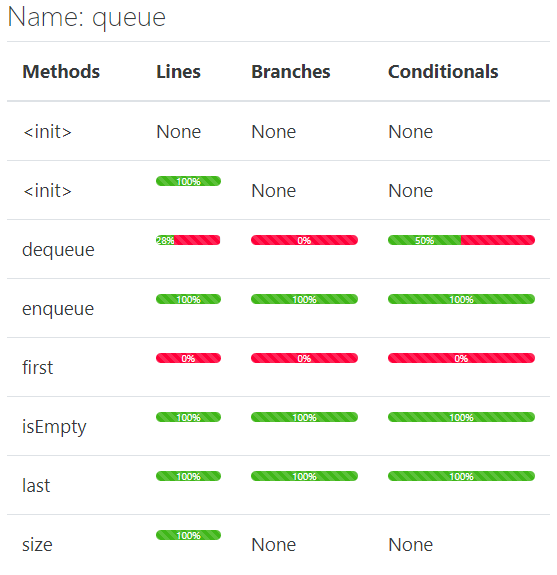}
  \caption{\textit{Detailed} feedback example}
  \label{fig:detailed-feedback}
\end{figure}

\begin{figure}[!htb]
  \centering
  \includegraphics[width=\linewidth]{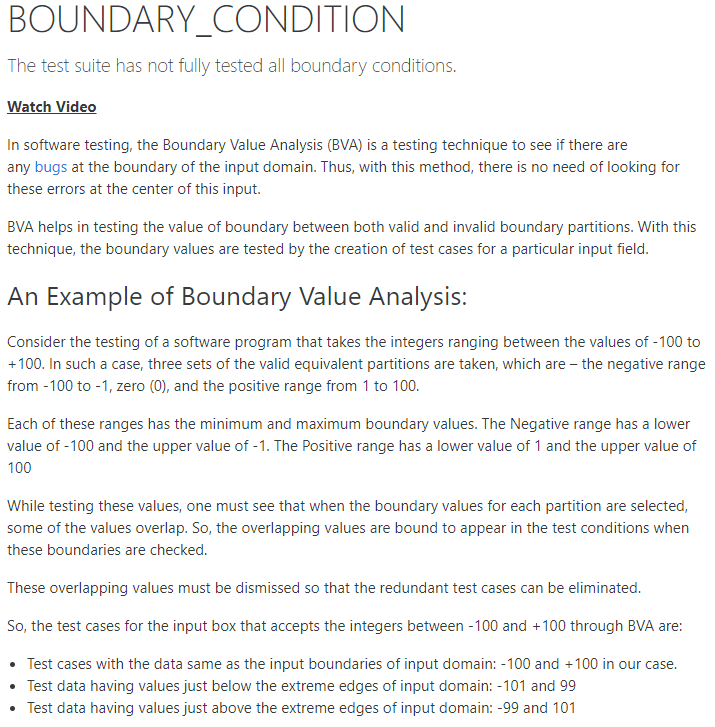}
  \caption{\textit{Conceptual} feedback example}
  \label{fig:conceptual-feedback}
\end{figure}

Figure~\ref{fig:detailed-feedback} provides an example of the \textit{detailed} coverage type of feedback (Treatment A). 
Figure~\ref{fig:conceptual-feedback} provides an annotated example of \textit{conceptual} feedback (Treatment B). 

\subsection{Study Design}
We received IRB approval for the studies. 
We employed the same seven phases in both instances of the course. 
When using Testing Tutor, we instructed the students to create the smallest, yet most complete test suite possible. 
We allowed the students to submit their code and tests as often as they would like to receive feedback.
\begin{itemize}
    \item \textit{Phase 1 - Training} : Training on using Testing Tutor.
    \item \textit{Phase 2 - Pre-test} : Students completed Assignment 1 using Testing Tutor, configured to provide no feedback (baseline).
    \item \textit{Phases 3-5 - Data Collection} : Students completed assignments 2-4 using Testing Tutor. 
    Students in Group A received the \textit{detailed} feedback (Treatment A).
    Students in Group B received the \textit{conceptual} feedback (Treatment B).  
    \item \textit{Session 6 - Post-test} : Students completed assignment 5 using Testing Tutor, configured to provide no feedback.
    This data helps us understand the students' retention of information learned from the feedback provided by Treatments A and B.
    \item \textit{Session 7 - Survey} : Students complete a survey about their experience using Testing Tutor and the likelihood they would use it in future courses. 
\end{itemize}

\subsection{Dependent Variables}
To help answer our research questions, we gathered data to support multiple dependent variables.
First, for each assignment submission, Testing Tutor collected the following data:
\begin{itemize}
    \item \textit{Line coverage} - The percentage of line coverage obtained.
    \item \textit{Branch coverage} - The percentage of branch coverage obtained.
    \item \textit{Conditional coverage} - The percentage of conditional coverage obtained.
    \item \textit{Redundant tests} - The number of tests in the test suite that are redundant (i.e. test code that is tested by another test).
\end{itemize}
Second, for each assignment we gathered the following data outside of Testing Tutor: 
\begin{itemize}
    \item \textit{Assignment grade} - The instructor assigns a grade for the quality of the test suite based on a rubric\footnote{\url{https://github.com/TestingTutor/Data/blob/master/SIGCSE21/Rubric.pdf}} that includes code coverage achieved, the number of redundant tests, and a visual inspection of test quality.
\end{itemize}
Lastly, we collected the following data at the end of the study:
\begin{itemize}
    \item {\itshape Perception of student understanding of the feedback} - An end-of-study optional and anonymous survey gathered the students’ perception of the feedback provided by Testing Tutor as well as the usability of Testing Tutor.
\end{itemize}

\section{Results}
\label{sec:results}
\input{sec-results}

\section{Discussion}
\label{sec:discussion}
The objective of these experiments were to compare the effects of inquiry-based \textit{conceptual} feedback with those of more traditional feedback mechanisms for software testing education. We now discuss insights and possible implications for software testing education as well as the limitations to these studies.

\subsection{Answers to Research Questions}
To summarize the results in the previous section, we provide an answer for each research question.

\vspace{8pt}
\noindent
\textbf{RQ1 - Effects of Feedback}: 
Perhaps of greatest practical significance, our results show that students, who began on an equal footing relative to testing knowledge and skill (based on our pre-test), achieved significantly different levels of code coverage, test redundancies, and programming grades based on the type of feedback they received. 
On average, students who received \textit{conceptual} feedback had higher code coverage (line, branch, and conditions), fewer redundant test cases, and higher programming grades compared with the students who received \textit{detailed} feedback. 
This increased performance occurred both while the students received the feedback (during Assignments 2-4) and once the feedback was removed (during Assignment 5).

The fact that the effect carried over to Assignment 5 indicates students who received \textit{conceptual} feedback were able to learn how to be better testers, which resulted in their better performance on Assignment 5.
Therefore, we can conclude that student obtained more long-term benefits from the \textit{conceptual} feedback than from the \textit{detailed} feedback. 

\vspace{8pt}
\noindent
\textbf{RQ2 - Student Perceptions}: 
The end-of-study survey results illustrated the students’ preference for \textit{conceptual} feedback. 
Students who received \textit{conceptual} feedback indicated that Testing Tutor helped them meet the objectives of the assignments (achieving higher code coverage and reducing redundant tests) in a more productive and effective way than students who received \textit{detailed} feedback. 
The survey results also indicated that \textit{conceptual} feedback had an effect on the students' perception of Testing Tutor's usability, ease of use, and whether they would recommend Testing Tutor to someone learning software testing.
All of these results were statistically significant in favor of the \textit{conceptual} feedback.

\subsection{Threats to Validity}
\textbf{Internal validity}: Sample size is the primary threat to internal validity. 
The sample sizes were constrained by the size of the class sections at the university during the time the studies were conducted. 
Although our pre-test showed no built-in bias in the grouping, we used ANOVAs and t-tests, which assume a normal distribution.

\vspace{4pt}
\textbf{Construct validity}:
Our definition of a high-quality test suite is the primary threat to construct validity.
We used code coverage metrics and redundancies in the test suite as our measure of high-quality.
It is possible that other measures of test-suite quality could produce different results.
However, we believe that our definition is valid and represents an important way to measure test suite quality.
Another potential threat would be how the grades were assigned to programs.
To reduce the chances of bias, the person who graded the assignments was not aware of the student's treatment group. 

\vspace{4pt}
\textbf{External validity}:
Because the student population in this academic program tend to have some professional programming experience, they may not be representative of other student populations.
To increase external validity, we chose a series of assignments that would be common for a course of this level. 
Another potential threat is that the length of the semesters in which the studies were conducted was different due to a shortened summer term. 
To mitigate this threat, we kept the length of the time given for the assignments consistent between the two studies. 
Furthermore, the populations in the two studies were similar in terms of prior knowledge as validated by the pretests.  

\section{Conclusion and Future Work}
\label{sec:conclusion}
These outcomes of this study can be explained by the power of IBL-based \textit{conceptual} feedback. 
IBL-based \textit{conceptual} feedback informs the student about the underlying fundamental testing concepts that she or he did not test in her or his test suite, rather than providing the student with code coverage analytical feedback. 
This IBL type of feedback allows the student to determine on her or his own how to improve the test suite. 
The feedback also has the positive side-effects of helping the student gain knowledge, experience, and reinforcement of fundamental testing concepts, which. ultimately, will make the student better tester in the long term.

Testing Tutor's approach falls well within the purview of best pedagogy practices regarding providing students with the information to reach their learning objectives, rather than simply following the traditional right/wrong dichotomy of traditional testing coverage feedback. 
Testing Tutor trains students to think about testing in a specific and logical manner while still allowing them the opportunity to use their critical thinking skills to solve complex problems. From a pedagogical perspective, the results indicate that Testing Tutor can be used as an efficient modality to both analyze and reinforce testing concepts taught as part of the CS curriculum.

We plan additional development work for Testing Tutor including additional student and class analysis for the instructor, developing a plug-in that allows a student to submit their tests through an Integrated Development Environment (IDE), and additional user experience improvements. 
In addition, we also plan to perform additional empirical studies with the following objectives: 1) improve the feedback mechanisms 2) understand the effectiveness of Testing Tutor’s feedback mechanisms at different levels of the curriculum 3) understand how Testing Tutor can be used as a tool for instructors to gauge learning and determine whether intervention is necessary to improve students’ learning.

\bibliographystyle{ACM-Reference-Format}
\bibliography{references} 

\end{document}

%% file: sec-results.tex
We organize this section around the study phases (Section~\ref{sec:Experiment}).
We do not have any \textit{a priori} reason to believe that the students in the two semesters are significantly different from each other.
Therefore, in the analyses that follow, we combine the data from both semesters.
In our ANOVA tests, we include the semester as one of the factors.
For the sake of space, we do not include all the details of the ANOVA results in the paper. 
These are available in the online appendix\footnote{\url{https://github.com/TestingTutor/Data/tree/master/SIGCSE21}}.

\subsection{Pre-test}
First, we analyzed the performance of the students on the pre-test (Assignment 1) to serve as a baseline and ensure no systematic differences between the groups.
Table~\ref{tab:pretest-results} provides an overview of the data.
The results of the 2-way ANOVA tests for each of the five dependent variables collected on Assignment 1 shows:
\begin{itemize}
    \item No significant differences between the groups for any of the dependent variables
    \item No main or interaction effects from the semester
\end{itemize}
Therefore, we can conclude based on the pre-test, that there was no built-in bias in the grouping.

\begin{table}[!htb]
  \caption{Pre-test Results}
  \label{tab:pretest-results}
  \input{table-pretest2}
\end{table}

\subsection{Comparison of Approaches}
Next, we analyzed whether there were any differences in the dependent variables between the two groups for Assignments 2-4. 
In this case, the ANOVA tests were slightly more complicated.
Because we had no reason to believe the specific assignments would impact the dependent variables, we analyzed Assignments 2-4 together.
As a result, we added a factor to the ANOVA.
For these five ANOVAs (one for each dependent variable), we have the following factors:
\begin{itemize}
    \item Treatment (Detailed/Conceptual)
    \item Assignment (2/3/4)
    \item Semester (Spring/Summer)
\end{itemize}
Table~\ref{tab:main-results} overviews the results which show:.
\begin{itemize}
    \item Significant differences (p $<$ .001) for all dependent variables;
    \item One case (Branch Coverage) in which the Assignment factor had showed a main effect;
    \item One case (Conditional Coverage) in which the Semester showed a main effect; and
    \item Two cases (Branch Coverage and Grade) where the Assignment showed an interaction with the Treatment
\end{itemize}
Overall, the results show students who received the \textit{conceptual} feedback performed significantly better than those who received the \textit{detailed} feedback.
At this point, we do not have a good explanation for few cases where other factors showed main or interaction effects, but will continue to explore these in future studies.

\begin{table}[!htb]
  \caption{Main Results}
  \label{tab:main-results}
  \input{table-mainstudy2}
\end{table}

\subsection{Post-test}
The results of the post-test (Assignment 5) provide insight into whether the type of feedback received on Assignments 2-4 affected the students' ability to test when that feedback was removed.
In Assignment 5, the students received no Testing Tutor feedback.
In this case, we ran five 2-way ANOVAs, similar to the pre-test.
The results of these tests (summarized in Table~\ref{tab:posttest-results}) showed:
\begin{itemize}
    \item Significant differences for all five dependent variables; and
    \item No main or interaction effects from the Semester
\end{itemize}
Therefore, we can conclude that the type of feedback received did have an effect on learning.
The students who received the \textit{conceptual} feedback were able to carry over what they learned more effectively than those received the \textit{detailed} feedback.

\begin{table}[!htb]
  \caption{Post-test Results}
  \label{tab:posttest-results}
  \input{table-posttest2}
\end{table}

\subsection{Survey}
The end-of-study optional survey contained nine questions each using a 7-point rating scale and three open-ended questions to gather information about usability. 
To examine whether there was any significant difference between the students in the two groups, we conducted a t-test for each question. 
Table~\ref{tab:survey-results} shows the averages for each group across both studies.
In all cases, those receiving \textit{conceptual} feedback viewed Testing Tutor significantly more favorably.
Due to space, detailed results are available in the online appendix.

\begin{table*}
  \caption{Survey Results (all questions on a 7-point scale)}
  \label{tab:survey-results}
  \input{table-surveyresults2}
\end{table*}

%% file: table-pretest2.tex
\begin{tabular}{M{.33\columnwidth}M{.25\columnwidth}M{.25\columnwidth}}
    \toprule 
    \textbf{Dependent Variable} & \textbf{Treatment A (Detailed)} & \textbf{Treatment B (Conceptual)} \\
    \midrule
    Line Coverage & 35\% & 35.7\% \\
    Branch Coverage & 35.3\% & 34.9\% \\
    Conditional Coverage & 35.1\% & 36.6\% \\
    Redundant Tests & 4.86 & 4.90 \\
    Assignment Grade & 57.95\% & 58.42\% \\
    \bottomrule
\end{tabular}

%% file: table-mainstudy2.tex
\begin{tabular}{M{.33\columnwidth}M{.25\columnwidth}M{.25\columnwidth}}
    \toprule 
    \textbf{Dependent Variable} & \textbf{Treatment A (Detailed)} & \textbf{Treatment B (Conceptual)} \\
    \midrule
    Line Coverage & 43.4\% & 55.1\% \\
    Branch Coverage & 43.1\% & 52.7\% \\
    Conditional Coverage & 45.4\% & 57.5\% \\
    Redundant Tests & 4.86 & 3.33 \\
    Assignment Grade & 60.37\% & 68.27\% \\
    \bottomrule
    \multicolumn{3}{l}{*[all differences significant p $<$ .05]}
\end{tabular}

%% file: table-posttest2.tex
\begin{tabular}{M{.33\columnwidth}M{.25\columnwidth}M{.25\columnwidth}}
    \toprule 
    \textbf{Dependent Variable} & \textbf{Treatment A (Detailed)} & \textbf{Treatment B (Conceptual)} \\
    \midrule
    Line Coverage & 37.9\% & 68.8\% \\
    Branch Coverage & 38.6\% & 69.4\% \\
    Conditional Coverage & 44.8\% & 72.6\% \\
    Redundant Tests & 4.29 & 2.29 \\
    Assignment Grade & 60.31\% & 78.95\% \\
    \bottomrule
    \multicolumn{3}{l}{*[all differences significant p $<$ .05]}
\end{tabular}

%% file: table-surveyresults2.tex
\begin{tabular}{M{.02\textwidth}p{.7\textwidth}|M{.1\textwidth}|M{.1\textwidth}}
\toprule
& Question Text & Treatment A (Detailed) & Treatment B (Conceptual) \\
\midrule
1 & The information that Testing Tutor provided helped me discover deficiencies in the code coverage. & 3.57 & 5.82  \\ 
\midrule
2 & The information Testing Tutor provided helped me discover redundant tests. & 4.25 & 5.21  \\
\midrule
3 & The information Testing Tutor provided regarding code coverage deficiencies made a lasting impression on how I approach software testing in the future. & 3.75 & 5.52  \\ 
\midrule 
4 & The information Testing Tutor provided regarding redundant tests made a lasting impression on how I approach software testing in the future. & 3.18 & 4.85  \\
\midrule
5 & Testing Tutor helped me become more EFFECTIVE at testing (achieving higher code coverage and reducing redundant tests) & 3.43 & 5.61 \\
\midrule
6 & Testing Tutor helped me become more PRODUCTIVE at testing (achieving higher code coverage and reducing redundant tests during the amount of time spent). & 3.82 & 5.97  \\
\midrule
7 & Testing Tutor is easy to use. & 3.89 & 5.79  \\
\midrule
8 & I learned to use Testing Tutor quickly. & 5.14 & 6.21  \\
\midrule
9 & I would recommend Testing Tutor to someone learning software testing. & 3.40 & 6.21  \\
\bottomrule
\multicolumn{4}{l}{*[all differences significant p $<$ .05]}
\end{tabular}

%% file: main.bbl

\begin{thebibliography}{29}


\ifx \showCODEN    \undefined \def \showCODEN     #1{\unskip}     \fi
\ifx \showDOI      \undefined \def \showDOI       #1{#1}\fi
\ifx \showISBNx    \undefined \def \showISBNx     #1{\unskip}     \fi
\ifx \showISBNxiii \undefined \def \showISBNxiii  #1{\unskip}     \fi
\ifx \showISSN     \undefined \def \showISSN      #1{\unskip}     \fi
\ifx \showLCCN     \undefined \def \showLCCN      #1{\unskip}     \fi
\ifx \shownote     \undefined \def \shownote      #1{#1}          \fi
\ifx \showarticletitle \undefined \def \showarticletitle #1{#1}   \fi
\ifx \showURL      \undefined \def \showURL       {\relax}        \fi
\providecommand\bibfield[2]{#2}
\providecommand\bibinfo[2]{#2}
\providecommand\natexlab[1]{#1}
\providecommand\showeprint[2][]{arXiv:#2}

\bibitem[\protect\citeauthoryear{Abell}{Abell}{2005}]%
        {abell_2005}
\bibfield{author}{\bibinfo{person}{Sandra~K. Abell}.}
  \bibinfo{year}{2005}\natexlab{}.
\newblock \showarticletitle{University Science Teachers as Researchers:
  Blurring the Scholarship Boundaries}.
\newblock \bibinfo{journal}{\emph{Research in Science Education}}
  \bibinfo{volume}{35}, \bibinfo{number}{2-3} (\bibinfo{year}{2005}),
  \bibinfo{pages}{281–298}.
\newblock
\urldef\tempurl%
\url{https://doi.org/10.1007/s11165-004-5600-x}
\showDOI{\tempurl}


\bibitem[\protect\citeauthoryear{Begel and Simon}{Begel and Simon}{2008}]%
        {begel_simon_2008}
\bibfield{author}{\bibinfo{person}{Andrew Begel} {and} \bibinfo{person}{Beth
  Simon}.} \bibinfo{year}{2008}\natexlab{}.
\newblock \showarticletitle{Struggles of new college graduates in their first
  software development job}.
\newblock \bibinfo{journal}{\emph{ACM SIGCSE Bulletin}} \bibinfo{volume}{40},
  \bibinfo{number}{1} (\bibinfo{year}{2008}), \bibinfo{pages}{226–230}.
\newblock
\urldef\tempurl%
\url{https://doi.org/10.1145/1352322.1352218}
\showDOI{\tempurl}


\bibitem[\protect\citeauthoryear{Buffardi and Edwards}{Buffardi and
  Edwards}{2015}]%
        {buffardi_2015}
\bibfield{author}{\bibinfo{person}{Kevin Buffardi} {and}
  \bibinfo{person}{Stephen~H. Edwards}.} \bibinfo{year}{2015}\natexlab{}.
\newblock \showarticletitle{Reconsidering Automated Feedback: A Test-Driven
  Approach}. In \bibinfo{booktitle}{\emph{Proceedings of the 46th ACM Technical
  Symposium on Computer Science Education}} (Kansas City, Missouri, USA)
  \emph{(\bibinfo{series}{SIGCSE '15})}. \bibinfo{publisher}{Association for
  Computing Machinery}, \bibinfo{address}{New York, NY, USA},
  \bibinfo{pages}{416–420}.
\newblock
\showISBNx{9781450329668}
\urldef\tempurl%
\url{https://doi.org/10.1145/2676723.2677313}
\showDOI{\tempurl}


\bibitem[\protect\citeauthoryear{Collofello and Vehathiri}{Collofello and
  Vehathiri}{2005}]%
        {collofello_vehathiri_2005}
\bibfield{author}{\bibinfo{person}{J. Collofello} {and} \bibinfo{person}{K.
  Vehathiri}.} \bibinfo{year}{2005}\natexlab{}.
\newblock \showarticletitle{An Environment for Training Computer Science
  Students on Software Testing}.
\newblock \bibinfo{journal}{\emph{Proceedings Frontiers in Education 35th
  Annual Conference}}.
\newblock
\urldef\tempurl%
\url{https://doi.org/10.1109/fie.2005.1611937}
\showDOI{\tempurl}


\bibitem[\protect\citeauthoryear{{de Souza}, {Oliveira}, {Maldonado}, {Souza},
  and {Barbosa}}{{de Souza} et~al\mbox{.}}{2014}]%
        {souza_2014}
\bibfield{author}{\bibinfo{person}{D.~M. {de Souza}}, \bibinfo{person}{B.~H.
  {Oliveira}}, \bibinfo{person}{J.~C. {Maldonado}}, \bibinfo{person}{S.~R.~S.
  {Souza}}, {and} \bibinfo{person}{E.~F. {Barbosa}}.}
  \bibinfo{year}{2014}\natexlab{}.
\newblock \showarticletitle{Towards the use of an automatic assessment system
  in the teaching of software testing}. In \bibinfo{booktitle}{\emph{2014 IEEE
  Frontiers in Education Conference (FIE) Proceedings}}. \bibinfo{pages}{1--8}.
\newblock


\bibitem[\protect\citeauthoryear{Edwards}{Edwards}{2003a}]%
        {edwards_2003}
\bibfield{author}{\bibinfo{person}{Stephen~H. Edwards}.}
  \bibinfo{year}{2003}\natexlab{a}.
\newblock \showarticletitle{Improving student performance by evaluating how
  well students test their own programs}.
\newblock \bibinfo{journal}{\emph{Journal on Educational Resources in
  Computing}} \bibinfo{volume}{3}, \bibinfo{number}{3} (\bibinfo{year}{2003}),
  \bibinfo{pages}{1–24}.
\newblock
\urldef\tempurl%
\url{https://doi.org/10.1145/1029994.1029995}
\showDOI{\tempurl}


\bibitem[\protect\citeauthoryear{Edwards}{Edwards}{2003b}]%
        {edwards_2003_1}
\bibfield{author}{\bibinfo{person}{Stephen~H. Edwards}.}
  \bibinfo{year}{2003}\natexlab{b}.
\newblock \showarticletitle{Teaching Software Testing: Automatic Grading Meets
  Test-First Coding}. In \bibinfo{booktitle}{\emph{Companion of the 18th Annual
  ACM SIGPLAN Conference on Object-Oriented Programming, Systems, Languages,
  and Applications}} (Anaheim, CA, USA) \emph{(\bibinfo{series}{OOPSLA '03})}.
  \bibinfo{publisher}{Association for Computing Machinery},
  \bibinfo{address}{New York, NY, USA}, \bibinfo{pages}{318–319}.
\newblock
\showISBNx{1581137516}
\urldef\tempurl%
\url{https://doi.org/10.1145/949344.949431}
\showDOI{\tempurl}


\bibitem[\protect\citeauthoryear{Edwards}{Edwards}{2004}]%
        {edwards_2004}
\bibfield{author}{\bibinfo{person}{Stephen~H. Edwards}.}
  \bibinfo{year}{2004}\natexlab{}.
\newblock \showarticletitle{Using software testing to move students from
  trial-and-error to reflection-in-action}.
\newblock \bibinfo{journal}{\emph{Proceedings of the 35th SIGCSE technical
  symposium on Computer science education - SIGCSE 04}} (\bibinfo{date}{May}
  \bibinfo{year}{2004}), \bibinfo{pages}{26}.
\newblock
\urldef\tempurl%
\url{https://doi.org/10.1145/971300.971312}
\showDOI{\tempurl}


\bibitem[\protect\citeauthoryear{Haddad}{Haddad}{2002}]%
        {haddad_2002}
\bibfield{author}{\bibinfo{person}{Hisham Haddad}.}
  \bibinfo{year}{2002}\natexlab{}.
\newblock \showarticletitle{Post-graduate assessment of CS students: experience
  and position paper}.
\newblock \bibinfo{journal}{\emph{Journal of Computing Sciences in Colleges}}
  \bibinfo{volume}{18}, \bibinfo{number}{2} (\bibinfo{year}{2002}),
  \bibinfo{pages}{189–197}.
\newblock


\bibitem[\protect\citeauthoryear{Houlden, Raja, Collier, Clark, and
  Waugh}{Houlden et~al\mbox{.}}{2004}]%
        {houlden_raja_collier_clark_waugh_2004}
\bibfield{author}{\bibinfo{person}{Robyn~L. Houlden},
  \bibinfo{person}{Jamila~B. Raja}, \bibinfo{person}{Christine~P. Collier},
  \bibinfo{person}{Albert~F. Clark}, {and} \bibinfo{person}{Jennifer~M.
  Waugh}.} \bibinfo{year}{2004}\natexlab{}.
\newblock \showarticletitle{Medical students’ perceptions of an undergraduate
  research elective}.
\newblock \bibinfo{journal}{\emph{Medical Teacher}} \bibinfo{volume}{26},
  \bibinfo{number}{7} (\bibinfo{year}{2004}), \bibinfo{pages}{659–661}.
\newblock
\urldef\tempurl%
\url{https://doi.org/10.1080/01421590400019542}
\showDOI{\tempurl}


\bibitem[\protect\citeauthoryear{Hu, Kussmaul, Knaeble, Mayfield, and Yadav}{Hu
  et~al\mbox{.}}{2016}]%
        {hu_kussmaul_kneable_mayfield_yadav_2016}
\bibfield{author}{\bibinfo{person}{Helen~H. Hu}, \bibinfo{person}{Clifton
  Kussmaul}, \bibinfo{person}{Brian Knaeble}, \bibinfo{person}{Chris Mayfield},
  {and} \bibinfo{person}{Aman Yadav}.} \bibinfo{year}{2016}\natexlab{}.
\newblock \showarticletitle{Results from a Survey of Faculty Adoption of
  Process Oriented Guided Inquiry Learning (POGIL) in Computer Science}. In
  \bibinfo{booktitle}{\emph{Proceedings of the 2016 ACM Conference on
  Innovation and Technology in Computer Science Education}} (Arequipa, Peru)
  \emph{(\bibinfo{series}{ITiCSE '16})}. \bibinfo{publisher}{Association for
  Computing Machinery}, \bibinfo{address}{New York, NY, USA},
  \bibinfo{pages}{186–191}.
\newblock
\showISBNx{9781450342315}
\urldef\tempurl%
\url{https://doi.org/10.1145/2899415.2899471}
\showDOI{\tempurl}


\bibitem[\protect\citeauthoryear{Justice, Warry, and Cuneo}{Justice
  et~al\mbox{.}}{2002}]%
        {justice_warry_cuneo_2002}
\bibfield{author}{\bibinfo{person}{Christopher Justice}, \bibinfo{person}{Wayne
  Warry}, {and} \bibinfo{person}{Carl Cuneo}.} \bibinfo{year}{2002}\natexlab{}.
\newblock \showarticletitle{A grammar for inquiry: linking goals and methods in
  a collaboratively taught social sciences inquiry course}.
\newblock \bibinfo{journal}{\emph{Alan Blizzard Award}} (\bibinfo{year}{2002}),
  \bibinfo{pages}{15–27}.
\newblock


\bibitem[\protect\citeauthoryear{Kazerouni, Shaffer, Edwards, and
  Servant}{Kazerouni et~al\mbox{.}}{2019}]%
        {kazerouni_2019}
\bibfield{author}{\bibinfo{person}{Ayaan~M. Kazerouni},
  \bibinfo{person}{Clifford~A. Shaffer}, \bibinfo{person}{Stephen~H. Edwards},
  {and} \bibinfo{person}{Francisco Servant}.} \bibinfo{year}{2019}\natexlab{}.
\newblock \showarticletitle{Assessing Incremental Testing Practices and Their
  Impact on Project Outcomes}. In \bibinfo{booktitle}{\emph{Proceedings of the
  50th ACM Technical Symposium on Computer Science Education}} (Minneapolis,
  MN, USA) \emph{(\bibinfo{series}{SIGCSE '19})}.
  \bibinfo{publisher}{Association for Computing Machinery},
  \bibinfo{address}{New York, NY, USA}, \bibinfo{pages}{407–413}.
\newblock
\showISBNx{9781450358903}
\urldef\tempurl%
\url{https://doi.org/10.1145/3287324.3287366}
\showDOI{\tempurl}


\bibitem[\protect\citeauthoryear{Kitchenham, Budgen, Brereton, and
  Woodall}{Kitchenham et~al\mbox{.}}{2005}]%
        {kitchenham_budgen_brereton_woodall_2005}
\bibfield{author}{\bibinfo{person}{Barbara Kitchenham}, \bibinfo{person}{David
  Budgen}, \bibinfo{person}{Pearl Brereton}, {and} \bibinfo{person}{Philip
  Woodall}.} \bibinfo{year}{2005}\natexlab{}.
\newblock \showarticletitle{An investigation of software engineering
  curricula}.
\newblock \bibinfo{journal}{\emph{Journal of Systems and Software}}
  \bibinfo{volume}{74}, \bibinfo{number}{3} (\bibinfo{year}{2005}),
  \bibinfo{pages}{325–335}.
\newblock


\bibitem[\protect\citeauthoryear{Lethbridge}{Lethbridge}{1998}]%
        {lethbridge_1998}
\bibfield{author}{\bibinfo{person}{Timothy~C. Lethbridge}.}
  \bibinfo{year}{1998}\natexlab{}.
\newblock \showarticletitle{A survey of the relevance of computer science and
  software engineering education}.
\newblock \bibinfo{journal}{\emph{Proceedings 11th Conference on Software
  Engineering Education}}, \bibinfo{pages}{56–66}.
\newblock
\urldef\tempurl%
\url{https://doi.org/10.1109/csee.1998.658300}
\showDOI{\tempurl}


\bibitem[\protect\citeauthoryear{Lethbridge}{Lethbridge}{2000a}]%
        {lethbridge_2000}
\bibfield{author}{\bibinfo{person}{Timothy~C. Lethbridge}.}
  \bibinfo{year}{2000}\natexlab{a}.
\newblock \showarticletitle{Priorities for the education and training of
  software engineers}.
\newblock \bibinfo{journal}{\emph{Journal of Systems and Software}}
  \bibinfo{volume}{53}, \bibinfo{number}{1} (\bibinfo{year}{2000}),
  \bibinfo{pages}{53–71}.
\newblock


\bibitem[\protect\citeauthoryear{Lethbridge}{Lethbridge}{2000b}]%
        {lethbridge_2000_1}
\bibfield{author}{\bibinfo{person}{Timothy~C. Lethbridge}.}
  \bibinfo{year}{2000}\natexlab{b}.
\newblock \showarticletitle{What knowledge is important to a software
  professional?}
\newblock \bibinfo{journal}{\emph{Computer (Long. Beach. Calif)}}
  \bibinfo{volume}{33}, \bibinfo{number}{5} (\bibinfo{year}{2000}),
  \bibinfo{pages}{44–50}.
\newblock
\urldef\tempurl%
\url{https://doi.org/10.1109/2.841783}
\showDOI{\tempurl}


\bibitem[\protect\citeauthoryear{Muukkonen, Lakkala, and Hakkarainen}{Muukkonen
  et~al\mbox{.}}{2005}]%
        {muukkonen_lakkala_hakkarainen_2005}
\bibfield{author}{\bibinfo{person}{Hanni Muukkonen}, \bibinfo{person}{Minna
  Lakkala}, {and} \bibinfo{person}{Kai Hakkarainen}.}
  \bibinfo{year}{2005}\natexlab{}.
\newblock \showarticletitle{Technology-Mediation and Tutoring: How Do They
  Shape Progressive Inquiry Discourse?}
\newblock \bibinfo{journal}{\emph{Journal of the Learning Sciences}}
  \bibinfo{volume}{14}, \bibinfo{number}{4} (\bibinfo{year}{2005}),
  \bibinfo{pages}{527–565}.
\newblock
\urldef\tempurl%
\url{https://doi.org/10.1207/s15327809jls1404_3}
\showDOI{\tempurl}


\bibitem[\protect\citeauthoryear{Pettit, Homer, Gee, Mengel, and
  Starbuck}{Pettit et~al\mbox{.}}{2015}]%
        {pettit_2015}
\bibfield{author}{\bibinfo{person}{Raymond Pettit}, \bibinfo{person}{John
  Homer}, \bibinfo{person}{Roger Gee}, \bibinfo{person}{Susan Mengel}, {and}
  \bibinfo{person}{Adam Starbuck}.} \bibinfo{year}{2015}\natexlab{}.
\newblock \showarticletitle{An Empirical Study of Iterative Improvement in
  Programming Assignments}. In \bibinfo{booktitle}{\emph{Proceedings of the
  46th ACM Technical Symposium on Computer Science Education}} (Kansas City,
  Missouri, USA) \emph{(\bibinfo{series}{SIGCSE '15})}.
  \bibinfo{publisher}{Association for Computing Machinery},
  \bibinfo{address}{New York, NY, USA}, \bibinfo{pages}{410–415}.
\newblock
\showISBNx{9781450329668}
\urldef\tempurl%
\url{https://doi.org/10.1145/2676723.2677279}
\showDOI{\tempurl}


\bibitem[\protect\citeauthoryear{Prince and Felder}{Prince and Felder}{2006}]%
        {prince_felder_2006}
\bibfield{author}{\bibinfo{person}{Michael~J. Prince} {and}
  \bibinfo{person}{Richard~M. Felder}.} \bibinfo{year}{2006}\natexlab{}.
\newblock \showarticletitle{Inductive Teaching and Learning Methods:
  Definitions, Comparisons, and Research Bases}.
\newblock \bibinfo{journal}{\emph{Journal of Engineering Education}}
  \bibinfo{volume}{95}, \bibinfo{number}{2} (\bibinfo{year}{2006}),
  \bibinfo{pages}{123–138}.
\newblock
\urldef\tempurl%
\url{https://doi.org/10.1002/j.2168-9830.2006.tb00884.x}
\showDOI{\tempurl}


\bibitem[\protect\citeauthoryear{Shymansky, Hedges, and Woodworth}{Shymansky
  et~al\mbox{.}}{1990}]%
        {shymansky_hedges_woodworth_1990}
\bibfield{author}{\bibinfo{person}{James~A. Shymansky},
  \bibinfo{person}{Larry~V. Hedges}, {and} \bibinfo{person}{George Woodworth}.}
  \bibinfo{year}{1990}\natexlab{}.
\newblock \showarticletitle{A reassessment of the effects of inquiry-based
  science curricula of the 60s on student performance}.
\newblock \bibinfo{journal}{\emph{Journal of Research in Science Teaching}}
  \bibinfo{volume}{27}, \bibinfo{number}{2} (\bibinfo{year}{1990}),
  \bibinfo{pages}{127–144}.
\newblock
\urldef\tempurl%
\url{https://doi.org/10.1002/tea.3660270205}
\showDOI{\tempurl}


\bibitem[\protect\citeauthoryear{Smith}{Smith}{1996}]%
        {smith_1996}
\bibfield{author}{\bibinfo{person}{Deborah~A. Smith}.}
  \bibinfo{year}{1996}\natexlab{}.
\newblock \bibinfo{booktitle}{\emph{A Meta-Analysis of Student Outcomes
  Attributable to the Teaching of Science as Inquiry as Compared to Traditional
  Methodology}}.
\newblock \bibinfo{publisher}{Temple University}.
\newblock


\bibitem[\protect\citeauthoryear{Spacco, Hovemeyer, Pugh, Emad, Hollingsworth,
  and Padua-Perez}{Spacco et~al\mbox{.}}{2006}]%
        {spacco_hovemeyer_pugh_emad_hollingsworth_padua-perez_2006}
\bibfield{author}{\bibinfo{person}{Jaime Spacco}, \bibinfo{person}{David
  Hovemeyer}, \bibinfo{person}{William Pugh}, \bibinfo{person}{Fawzi Emad},
  \bibinfo{person}{Jeffrey~K. Hollingsworth}, {and} \bibinfo{person}{Nelson
  Padua-Perez}.} \bibinfo{year}{2006}\natexlab{}.
\newblock \showarticletitle{Experiences with marmoset}.
\newblock \bibinfo{journal}{\emph{Proceedings of the 11th annual SIGCSE
  conference on Innovation and technology in computer science education -
  ITICSE 06}} (\bibinfo{year}{2006}).
\newblock
\urldef\tempurl%
\url{https://doi.org/10.1145/1140124.1140131}
\showDOI{\tempurl}


\bibitem[\protect\citeauthoryear{Spronken-Smith, Angelo, Matthews, O'Steen, and
  Robertson}{Spronken-Smith et~al\mbox{.}}{2007}]%
        {spronken-smith_angelo_matthews_o'steen_robertson_2007}
\bibfield{author}{\bibinfo{person}{Rachel Spronken-Smith}, \bibinfo{person}{Tom
  Angelo}, \bibinfo{person}{Helen Matthews}, \bibinfo{person}{Billy O'Steen},
  {and} \bibinfo{person}{Jane Robertson}.} \bibinfo{year}{2007}\natexlab{}.
\newblock \showarticletitle{How Effective is Inquiry-Based Learning in Linking
  Teaching and Research?}
\newblock \bibinfo{journal}{\emph{An International Colloquium on International
  Policies and Practices for Academic Enquiry}} \bibinfo{volume}{7},
  \bibinfo{number}{4} (\bibinfo{year}{2007}), \bibinfo{pages}{1–7}.
\newblock


\bibitem[\protect\citeauthoryear{Vygotskij and Cole}{Vygotskij and
  Cole}{1981}]%
        {vygotskij_cole_1981}
\bibfield{author}{\bibinfo{person}{Lev~S. Vygotskij} {and}
  \bibinfo{person}{Michael Cole}.} \bibinfo{year}{1981}\natexlab{}.
\newblock \bibinfo{booktitle}{\emph{Mind in society: the development of higher
  psychological processes}}.
\newblock \bibinfo{publisher}{Cambridge: Harvard University Press}.
\newblock


\bibitem[\protect\citeauthoryear{Wang, Schwartz, and Lingard}{Wang
  et~al\mbox{.}}{2008}]%
        {wang_2008}
\bibfield{author}{\bibinfo{person}{Taehyung~(George) Wang},
  \bibinfo{person}{Diane Schwartz}, {and} \bibinfo{person}{Robert Lingard}.}
  \bibinfo{year}{2008}\natexlab{}.
\newblock \showarticletitle{Assessing Student Learning in Software
  Engineering}.
\newblock \bibinfo{journal}{\emph{J. Comput. Sci. Coll.}} \bibinfo{volume}{23},
  \bibinfo{number}{6} (\bibinfo{date}{June} \bibinfo{year}{2008}),
  \bibinfo{pages}{239–248}.
\newblock
\showISSN{1937-4771}
\urldef\tempurl%
\url{https://doi.org/10.5555/1352383.1352424}
\showDOI{\tempurl}


\bibitem[\protect\citeauthoryear{Yarger, Gallus, Taber, Boysen, and
  Castleberry}{Yarger et~al\mbox{.}}{2000}]%
        {yarger_gallus_taber_boysen_castleberry_2000}
\bibfield{author}{\bibinfo{person}{Douglas~N. Yarger},
  \bibinfo{person}{William~A. Gallus}, \bibinfo{person}{Michael Taber},
  \bibinfo{person}{J.~Peter Boysen}, {and} \bibinfo{person}{Paul Castleberry}.}
  \bibinfo{year}{2000}\natexlab{}.
\newblock \showarticletitle{A Forecasting Activity for a Large Introductory
  Meteorology Course}.
\newblock \bibinfo{journal}{\emph{Bulletin of the American Meteorological
  Society}} \bibinfo{volume}{81}, \bibinfo{number}{1} (\bibinfo{year}{2000}),
  \bibinfo{pages}{31–39}.
\newblock
\urldef\tempurl%
\url{https://doi.org/10.1175/1520-0477(2000)081<0031:afafal>2.3.co;2}
\showDOI{\tempurl}


\bibitem[\protect\citeauthoryear{Yin}{Yin}{2006}]%
        {Yin2006PreparingRA}
\bibfield{author}{\bibinfo{person}{R. Yin}.} \bibinfo{year}{2006}\natexlab{}.
\newblock \showarticletitle{Preparing Resource and Environmental Managers with
  International Understanding and Merits (PREMIUM): Introducing a research
  experience for undergraduates program}.
\newblock \bibinfo{journal}{\emph{Journal of Forestry}}  \bibinfo{volume}{104}
  (\bibinfo{year}{2006}), \bibinfo{pages}{320--323}.
\newblock


\bibitem[\protect\citeauthoryear{Zoller}{Zoller}{1999}]%
        {zoller_1999}
\bibfield{author}{\bibinfo{person}{Uri Zoller}.}
  \bibinfo{year}{1999}\natexlab{}.
\newblock \showarticletitle{Scaling-up of higher-order cognitive
  skills-oriented college chemistry teaching: An action-oriented research}.
\newblock \bibinfo{journal}{\emph{Journal of Research in Science Teaching}}
  \bibinfo{volume}{36}, \bibinfo{number}{5} (\bibinfo{year}{1999}),
  \bibinfo{pages}{583--596}.
\newblock
\urldef\tempurl%
\url{https://doi.org/10.1002/(SICI)1098-2736(199905)36:5<583::AID-TEA5>3.0.CO;2-M}
\showDOI{\tempurl}


\end{thebibliography}
